%% file: nomen.tex
\documentclass{amsart}
\makeatletter
\def\ps@headings{%
\def\@oddhead{\mbox{}\scriptsize\rightmark \hfil \thepage}%
\def\@evenhead{\scriptsize\thepage \hfil \leftmark\mbox{}}%
\def\@oddfoot{}%
\def\@evenfoot{}}
\makeatother
\usepackage{flushend}

\usepackage{graphicx, subfigure}
\graphicspath{{Fig/}}
\usepackage{color}
\usepackage{epsfig}

\usepackage{bbm}

\usepackage[latin1]{inputenc}

\def\etal{{\em et al. }}

\newcommand\ind[1]{\mathbbm{1}{\left\{#1\right\}}}

\title[Impact of traffic mix on caching performance]{Impact of traffic mix on caching performance in a content-centric network}

\author[C. Fricker]{Christine Fricker}

\author[Ph. Robert]{Philippe Robert}

\author[J. Roberts]{James Roberts}

\author[N. Sbihi]{Nada Sbihi}

\address[C. Fricker, Ph. Robert, J. Roberts, N. Sbihi]{INRIA Paris --- Rocquencourt,  Domaine de Voluceau, 78153 Le Chesnay, France.}

\email{Christine Fricker@inria.fr}

\email{Philippe.Robert@inria.fr}
\urladdr{http://www-rocq.inria.fr/\string~robert}

\email{James.Roberts@inria.fr}

\email{Nada.Sbihi@inria.fr}

\begin{document}
\begin{abstract}
For a realistic traffic mix, we evaluate the hit rates attained in a two-layer cache hierarchy designed to reduce Internet bandwidth requirements. The model identifies four main types of content, web, file sharing, user generated content and video on demand, distinguished in terms of their traffic shares, their population and object sizes and their popularity distributions. Results demonstrate that caching VoD in access routers offers a highly favorable bandwidth memory tradeoff but that the other types of content would likely be more efficiently handled in very large capacity storage devices in the core. Evaluations are based on a simple approximation for LRU cache performance that proves highly accurate in relevant configurations.

\end{abstract}

\maketitle


\section{Introduction}

With some 96\% of Internet traffic currently generated by users retrieving content of one form or another \cite{CVNI}, it is increasingly important to understand the memory bandwidth tradeoff achievable through caching, whether this be performed in content distribution networks overlaid on IP or in radical new Internet architectures like CCN \cite{JSTP09}. In this paper we investigate this tradeoff for a realistic traffic mix resulting from the web, file sharing and video content retrieval.

\begin{figure}[h]
\centering
\resizebox{5cm}{!}{\input{network.pdf_t}}

\caption{Considered two-layer cache hierarchy}
\label{fig:network}
\end{figure}
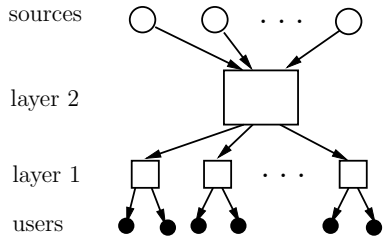

We consider a simple generic hierarchical network, as depicted in Figure \ref{fig:network}. A lower layer of caches, close to users, would be located in access routers and consists therefore of a large number of similarly sized content stores. The second layer would typically consist of a set of coordinated storage facilities located within the network core. These facilities might themselves be arranged in a hierarchy but, for present purposes, we assume they can be assimilated to a single large cache. All requests for content not satisfied at layer 1 are routed to layer 2. If the content cannot be found in either layer, requests are forwarded to a source situated outside the considered network. We seek to quantify bandwidth savings between the two layers, measuring in-network capacity gains, and beyond the second layer, measuring the reduction in traffic coming from sources, peers and transit providers. 

We suppose caching is performed at the level of entire named objects rather than chunks. As we show in the paper, performance does not depend critically on this assumption as long as all chunks of a given object are equally popular. We assume request processes follow the so-called ``independent reference model'' and mainly assume caches implement a ``least recently used'' (LRU) replacement policy. In some cases we also consider the potential gain from the more complex but hit rate optimal ``least frequently used'' (LFU) policy. 

Our main contribution is to evaluate the performance of this cache hierarchy under a demand model that reflects a realistic traffic mix. This model accounts for significant differences between the nature of web, file sharing, user generated content (UGC) and video on demand (VoD) content. These differences are manifested  through relative traffic proportions, object and population sizes and popularity distributions. We make extensive use of an approximation for LRU due initially to Che \etal \cite{CTW02} that is shown to be extremely accurate for a wider range of object populations and popularity distributions than previously thought. 

We are aware of the extensive prior work on cache performance, first  for computer memory management (e.g., \cite{FGT92, Jelenkovic99}), then for the web (e.g., \cite{Williamson02, LSS04}) and, more recently, for proposed content-centric network architectures (e.g., \cite{Psaras2011,CGM11}). It is unfortunately not possible in this short paper to summarize this and its relation to our own work. For the sake of clarity, the paper recalls some properties of caching that are already well-known from this prior work. 

We proceed in the next section by describing the characteristics of the four distinct types of Internet content retrieval traffic. In the following section, Section \ref{sec:homogeneous}, we discuss cache performance for homogeneous traffic and introduce the Che approximation. Section \ref{sec:mix} presents the results of applying this approach to evaluating cache performance under the considered realistic traffic mix. 

%


%

%


\section{Internet content characteristics}
\label{sec:traffic}

We discuss characteristics of Internet traffic that are significant for cache performance and deduce rough estimates of relevant parameter values.

\subsection{Types of content}
The Cisco Visual Networking Index published in 2011 classifies Internet traffic and forecasts global demand for the period 2010-2015 \cite{CVNI}. Some 96\% of traffic is content retrieval, classified as web, file sharing or video. We further divide video into user generated content (UGC) and video on demand (VoD), supposing equal volumes in 2011 and an increased proportion of VoD in 2015. Estimated traffic shares are listed in Table \ref{tab:characteristics}. 

\begin{table}[t]
\begin{center}
\begin{tabular}{l  | c c |c| c| c }
\hline
 & \multicolumn{2}{c|}{traffic share ($p_i$) } &  population  & mean object & overall \\
  & 2011 & 2015 & size ($N_i$)& size ($\theta_i$) & volume\\
   \hline
Web   & .18 & .16& $10^{11}$ & 10 KB & 1 PB\\
File sharing  & .36 & .24& $10^{5}$  &  10 GB & 1 PB\\
UGC  & .23 & .23&  $10^8$  &  10 MB & 1 PB \\
VoD & .23 & .37 & $10^4$ &  100 MB & 1 TB \\ 
\hline
\end{tabular}
\caption{Assumed characteristics of Internet content traffic}
\label{tab:characteristics}
\end{center}
\vspace{-10mm}
\end{table}

\subsection{Population size and object size}

In 2008 Google identified some $10^{12}$ unique URLs\footnote{http://www.boutell.com/newfaq/misc/sizeofweb.html}. We conservatively assume a total of $10^{11}$ distinctly named web elements and suppose these have mean size 10 KB  \cite{Mahanti00}.



To estimate the characteristics of file sharing content we use statistics derived from the BitTorrent tracker site Demonoid\footnote{www.demonoid.me/}. The site distinguishes more than $400\;000$ torrents. We compiled statistics on a representative sample to deduce a mean size of 7.4 GB (the largest torrents correspond to entire seasons of TV series). 

UGC content is dominated by YouTube. A recent study by Zhou \etal estimates there are currently $5 \times 10^8$ YouTube videos of mean size 10 MB  \cite{Zhou11}. VoD catalogues, on the other hand, are much smaller. Inspection of various sites yields populations measured in thousands of movies, TV shows and trailers. We estimate mean VoD object size to be around 100 MB.

Population and size estimates in Table \ref{tab:characteristics} are order of magnitude approximations derived from the above discussion. Note that it matters little for the present evaluations that objects be identified as such or as a larger set of smaller chunks. The most significant statistic is the overall volume: the first three types each count for 1 petabyte, three orders of magnitude greater than the assumed 1 terabyte of VoD content.

\subsection {Popularity distribution}
Caching performance  depends crucially on the relative popularity of different objects. It has frequently been observed that the popularity of web pages follows a generalized Zipf law: the request rate $q(n)$ for the $n^{th}$ most popular page is proportional to $1/n^\alpha$ for some $\alpha$. Measurements reported in \cite{Breslau99} and \cite{Mahanti00}, for instance, yield estimates for $\alpha$ between .64 and .83.

\begin{figure}[t]
\vspace{-7mm}
\centering
\includegraphics[scale=.64]{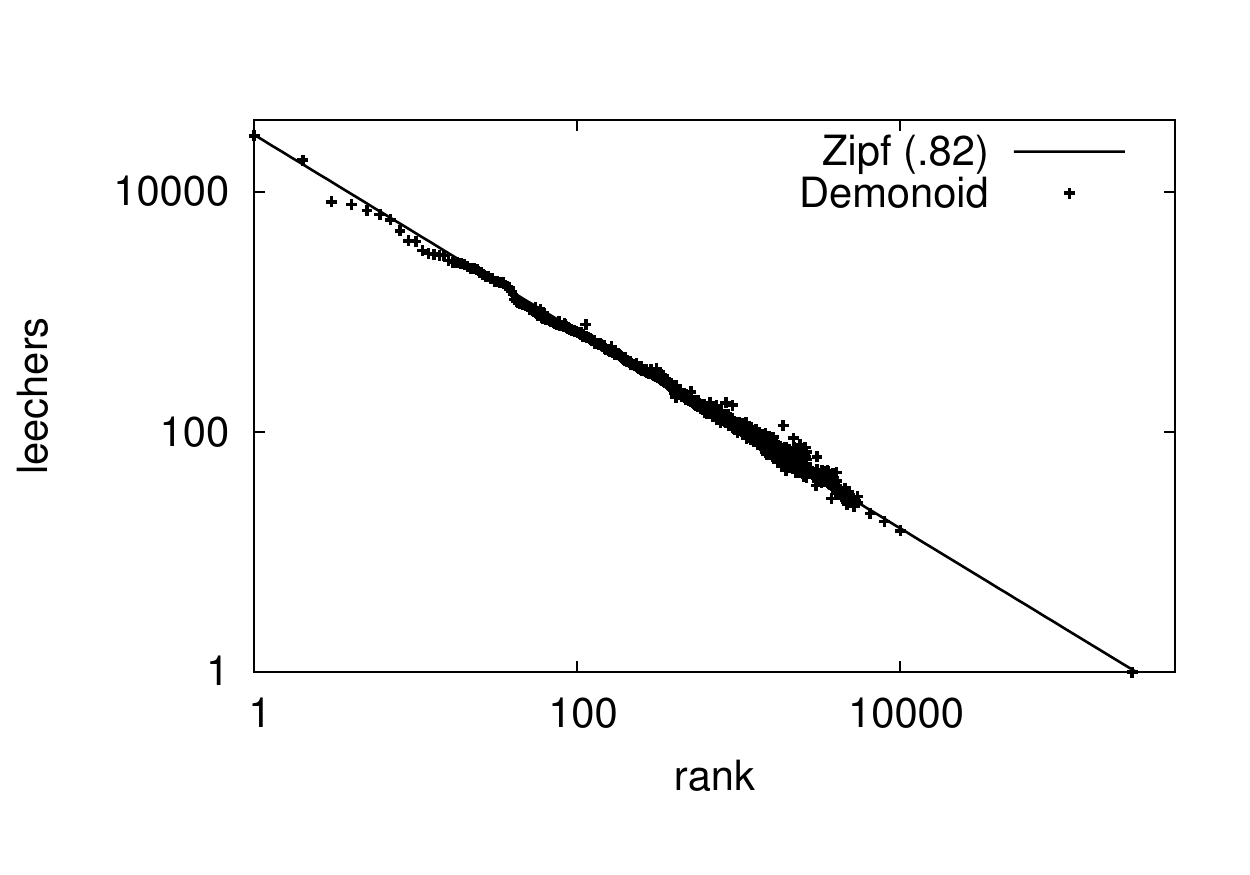}
\vspace{-6mm}
\caption{Popularity v rank for 270000 torrents on Demonoid.me}
\label{fig:demonoid}
\vspace{-3mm}
\end{figure}

Statistics on the Demonoid site allow us to measure the popularity of torrents. After entering a keyword, it is possible to sort torrents in decreasing order of the current number of leechers. This number is an accurate measure of instantaneous popularity since, by definition, a leecher is actively downloading chunks of the torrent. The keyword ``a'' selected $270\;000$ of the $400\;000$ total. The site in fact only displays statistics for the first and last $10\;000$ torrents. The popularity distribution displayed in Figure \ref{fig:demonoid} was derived from the former while the latter all had zero leechers. The data closely matches a Zipf law with $\alpha=.82$\footnote{The fact that the plot is not strictly decreasing arises from the delay between the sort and the moment when we read the statistics during which time the number of leechers changed.}. A similar capture for Pirate Bay also revealed Zipf law popularity with a smaller exponent, $\alpha=.75$. 

A Zipf law also appears to accurately characterize UGC popularity. The Zipf exponent was estimated at .56 by Gill \etal \cite{Gill07} while a higher value of around 0.8 can be deduced from the results of Cha \etal  \cite{Cha07}. Results in a recent technical report by Carlinet \etal suggest content of the UGC service DailyMotion has Zipf popularity with $ \alpha \approx .88$  \cite{Carlinet11} .

The study by Carlinet \etal also evaluates the popularity of a VoD service. Here the popularity law is not Zipf but might be approximated by a curve with two components: the first is relatively flat for the 100 most popular objects ($\sim$ Zipf(.5)) while the tail between  ranks 100 and 4000 is much steeper ($\sim$ Zipf(1.2)). Statistics gathered by Yu \etal for a VoD service in China suggest Zipf law popularity with $.65 \le \alpha \le 1$ \cite{Yu06}.

Many qualifying remarks could be made about the nature of the measurements used to derive the above estimates and one should take account of phenomena like spatial and temporal locality. However,  for the present investigation we will simply suppose web, file sharing and UGC content have Zipf law popularity with an exponent of 0.8. VoD demand is less well characterized. In view of its growing importance in the Internet traffic mix, we evaluate performance for two contrasting laws: Zipf(.8) like the other types, and Zipf(1.2), reflecting a possibly more accentuated popularity distribution.

\section{Caching homogeneous content}
\label{sec:homogeneous} 
We recall properties of caching when content objects have constant size and follow a Zipf popularity distribution.  An approximation for the LRU hit rate is described and used to evaluate the hit rate. Like prior work, we adopt the \emph{independent reference model}: the probability a content request is for a given object depends only on that object's popularity and not on the sequence of requests that came before.  Equivalently, we assume object requests occur at the instants of independent Poisson processes.  Some notation used in this and the next section is listed in Table \ref{tab:notation}.

\begin{table}[t]
\begin{center}
\begin{tabular}{l  l }
\hline
$C$  & cache size in objects or bytes \\
$p_i$ &  proportion of traffic of type $i$ \\
$N_i$   & number of objects of type $i$ \\
$\theta_i$  & mean object size for type $i$ \\
$\alpha_i$ & Zipf popularity law exponent for type $i$   \\
$q_i(n)$ & popularity (request rate) for object $n$ of type $i$  \\ 
$h_i(n)$ & hit rate for object $n$ of type $i$ \\
\hline
\end{tabular}
\caption{Principal notation}
\label{tab:notation}
\end{center}
\vspace{-10mm}

\end{table}

\subsection{LFU performance}
Under the independent reference model, LFU is known to maximize the hit rate and therefore constitutes a useful benchmark. It is also worth noting that performance close to that of LFU can be attained relatively simply using the notion of persistent access caching introduced by Jelenkovic and co-authors  \cite{JR08, JKR05}. With LFU, the hit rate for a cache of size $C$ is simply $\sum_{1\le n \le C} q(n)/ \sum_{1\le n \le N} q(n)$. 

\vspace{2mm}

\begin{figure}[t]
\centering
\includegraphics[scale=.6]{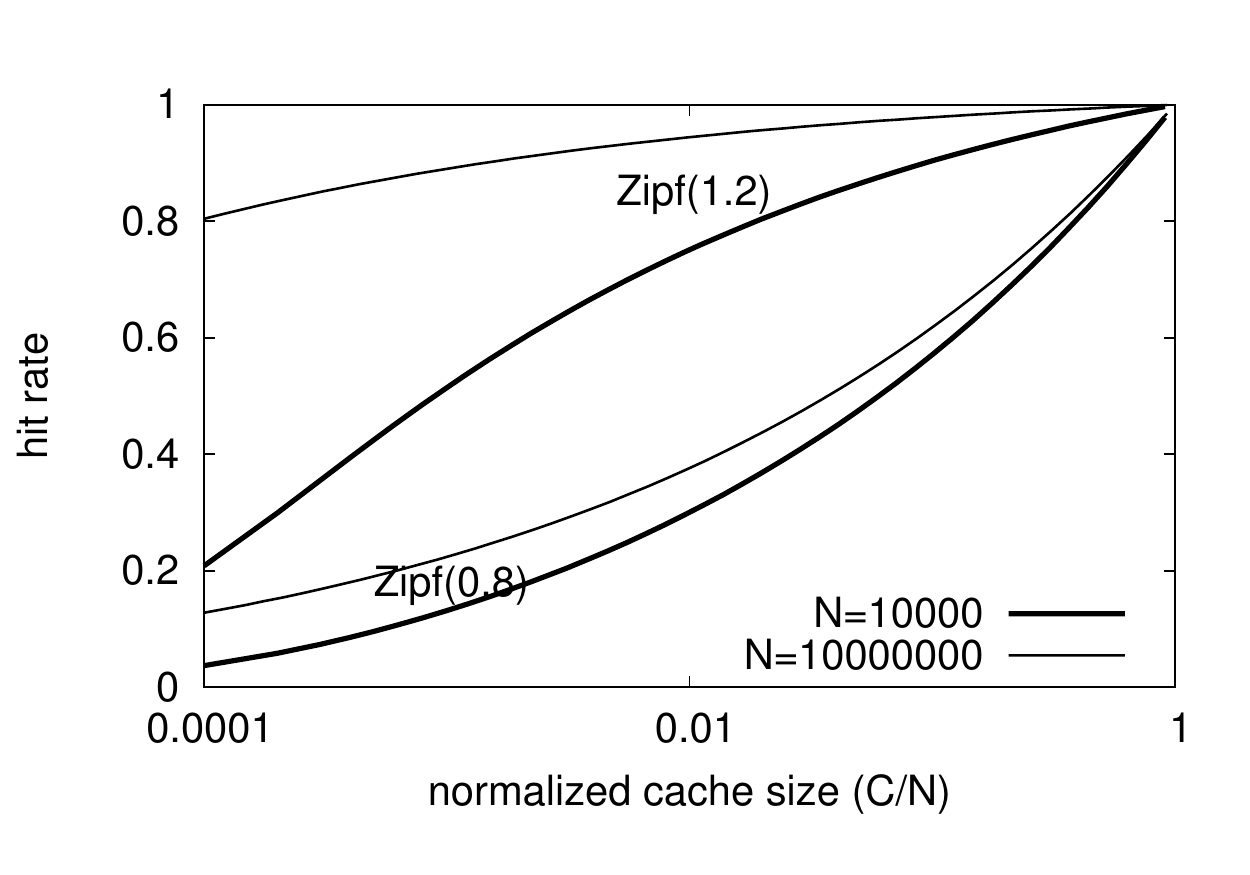}
\vspace{-4mm}
\caption{LFU hit rate v relative cache size: Zipf(.8) and Zipf(1.2) popularity.}
\label{fig:LFU}
\vspace{-3mm}

\end{figure}

Figure \ref{fig:LFU} plots LFU hit rate  against normalized cache size, $C/N$, for two Zipf parameters, $\alpha=.8$ and $\alpha=1.2$, and two populations, $N=10^4$ and $N=10^8$. The figure illustrates the qualitative difference in behaviour for $\alpha < 1$ and $\alpha>1$. In the former case, the hit rate as a function of $C/N$ tends to a limit as $N$ increases whereas, when $\alpha>1$, it is the hit rate as a function of $C$ that tends to a limit (note that this convergence cannot be seen on this figure). Caching is significantly less effective when $\alpha<1$. The population $N$ has a non-negligible impact on performance.

\subsection{LRU performance}
\label{sec:che}
To evaluate hit rates for LRU, we use an approximation proposed by Che \etal \cite{CTW02} that proves to be accurate for a wide variety of popularity distributions and population sizes. 

The Che approximation consists in assuming the hit rate $h(n)$ of an object $n$ with popularity $q(n)$ can be written $1- e^{-q(n) T_C}$ for some parameter $T_C$.  Since $C=\sum_n \ind{\textrm{cache contains object }n}$, taking expectations gives: 
\begin{equation}
C=\sum_n h(n) =\sum_n(1-e^{-q(n) T_C}).
\label{eq:che}
\end{equation}
 Solving (\ref{eq:che}) for $T_C$ yields the hit rates $h(n)$.

This approximation was introduced differently in \cite{CTW02}. Che and co-authors assimilate $T_C$ to the time needed for $C$ distinct objects to be requested assuming independent Poisson processes of requests for objects $n$ at rates $q(n)$. They argue that the approximation is accurate for large populations and large cache sizes since $T_C$ is then nearly deterministic. In this case, $1-e^{-q(n) T_C}$ corresponds to the object $n$ hit rate since it gives the probability the inter-request time is less than $T_C$.
In fact, our evaluations reveal that the approximation is accurate for a wide range of population sizes and popularity distributions where this intuitive argument is not always justified. 

\begin{figure}[t]
   \vspace{-4mm}
        \subfigure{%
            \label{fig:Zipf}
\hspace{-12mm}
            \includegraphics[width=0.35\textwidth]{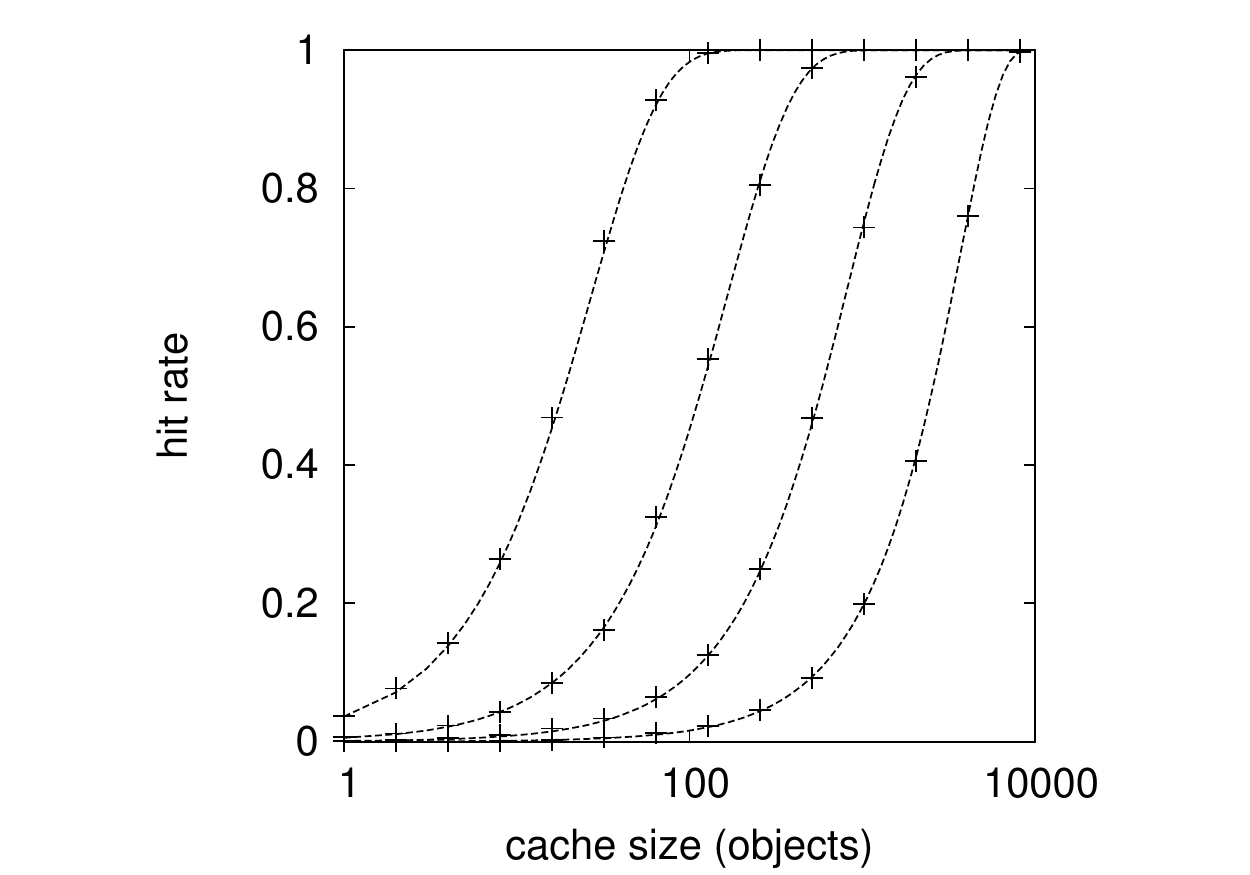}
        }%
        \subfigure{%
           \label{fig:geo}
\hspace{-22mm}
           \includegraphics[width=0.35\textwidth]{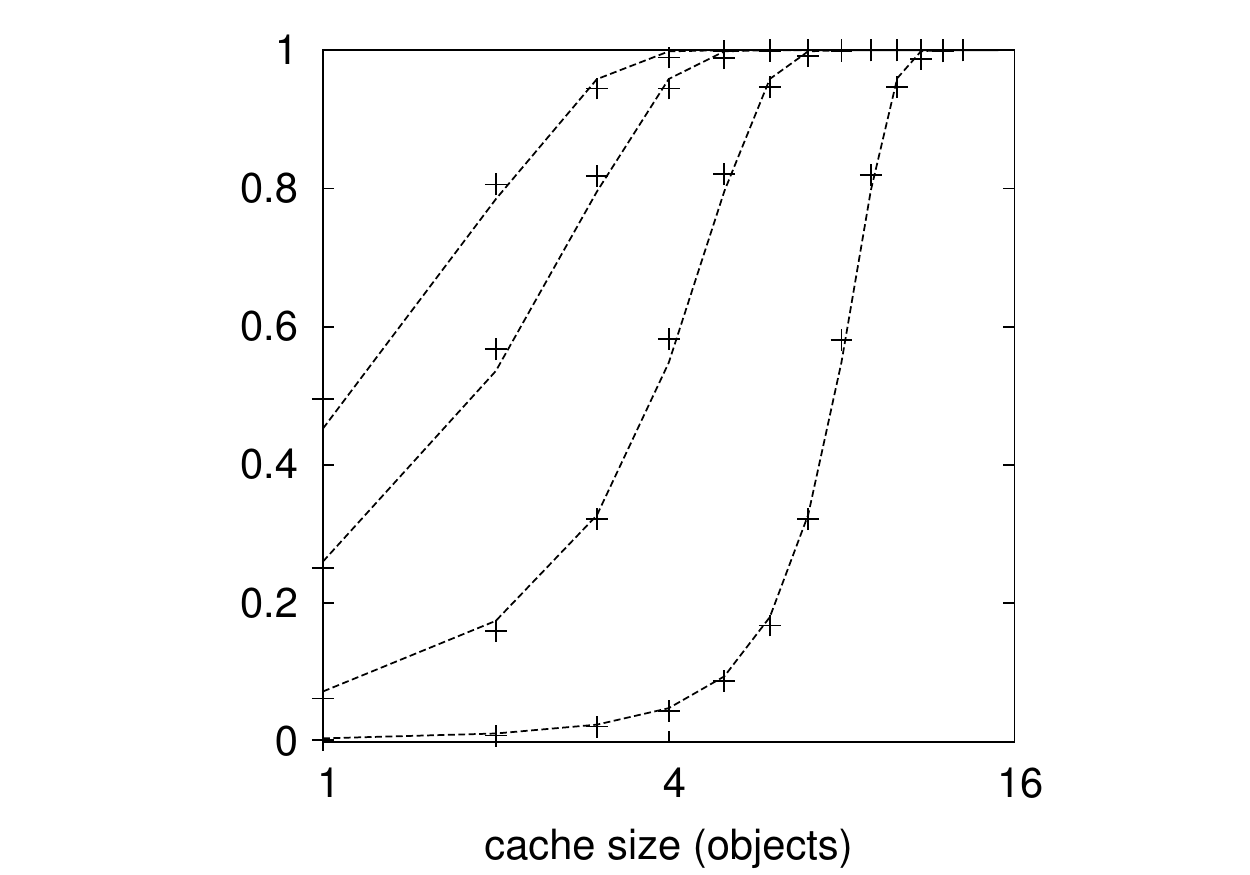}
        }   
    \caption{%
        Hit rate v cache size, crosses from simulation, lines from Che approximation: left, $N=10^4$, Zipf(.8) popularity, ranks 1, 10, 100, 1000; right, $N=16$, geo(.5) popularity, ranks 1, 2, 4, 8.
     }%
     \label{fig:haoche}
     \vspace{-4mm}

\end{figure}

The accuracy of the approximation is typified by the results shown in Figure \ref{fig:haoche}.   The plot on the left shows hit rates as a function of cache size for objects of popularity rank 1, 10, 100 and 1000 from a population of 10000 objects with Zipf(.8) popularity.  The right hand plot confirms the approach is accurate also for a small population of 16 objects with geometric  popularity, $q(n)=1/2^n$, where the intuitive arguments clearly do not apply. The method can readily be shown to be exact when the popularity distribution is uniform.

\begin{figure}[t]
\vspace{-3mm}

\centering
\includegraphics[scale=.6]{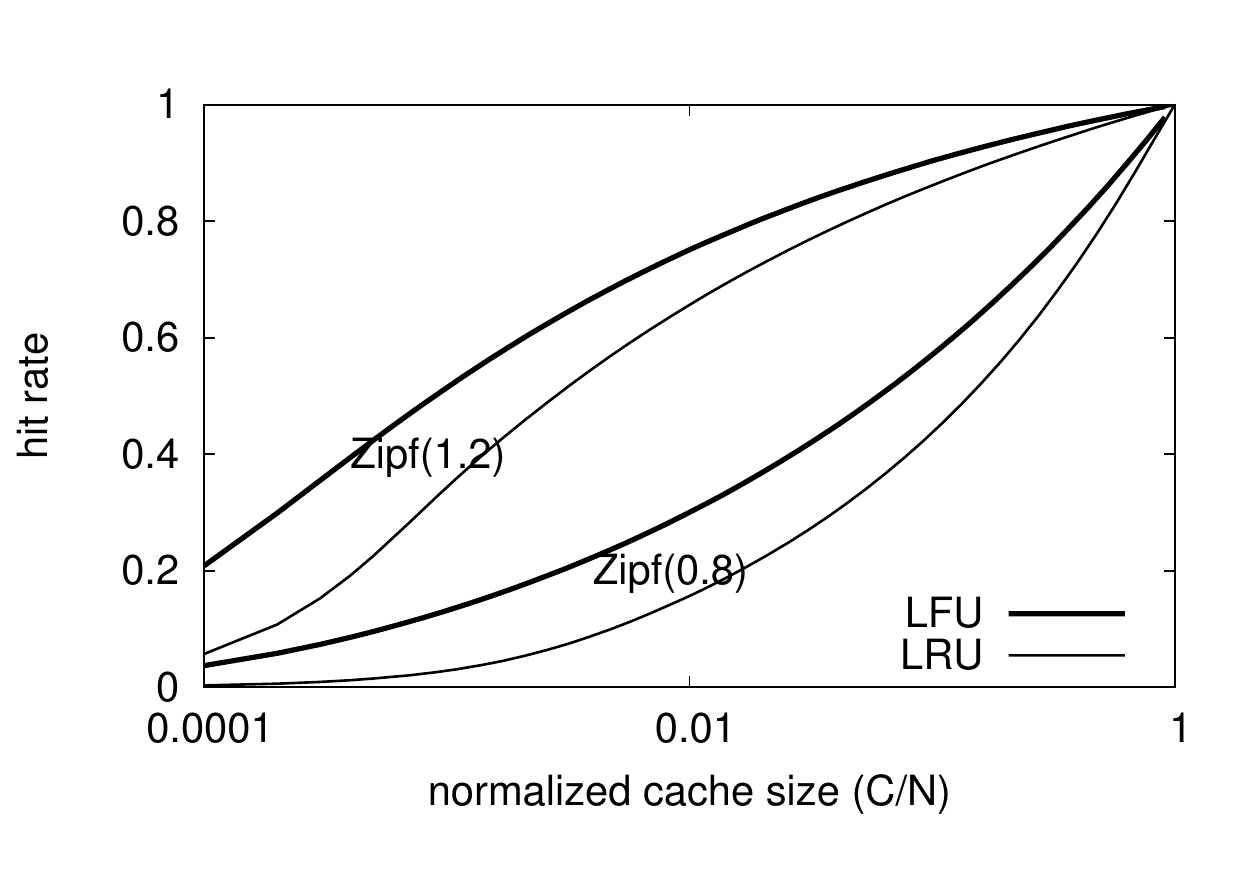}
\vspace{-4mm}
\caption{LFU and LRU hit rates v relative cache size: Zipf(.8) and Zipf(1.2) popularity, $N = 10^4$.}
\label{fig:LFU-LRU}
\end{figure}

Figure \ref{fig:LFU-LRU} compares the performance of LFU and LRU replacement policies for a population of $10^4$ objects with Zipf(.8) and Zipf(1.2) popularity distributions\footnote{The figure shows simulation results for LRU though the Che approximation yields  precisely the same plots.}. These results demonstrate the relative inefficiency of LRU compared to the optimal policy. This is particularly significant for small to medium size caches (up to 10\% of population, say) and the Zipf(.8) popularity distribution.

\subsection{A two-layer hierarchy}
Given the assumed large number of layer 1 caches, it is reasonable to apply the independent reference model at the second layer since its request process results from the superposition of many independent overflow processes, each contributing a small fraction of overall demand. Moreover, in this configuration the occupancy states of first and second layer caches can reasonably be assumed to be statistically independent.  

Suppose the size of each first layer cache is $C_1$ and the size of the second layer cache is $C_2$. Let $h(n)$ and $h'(n)$ denote the hit rate  for object $n$ at layers 1 and 2, respectively. We estimate $h(n)$ using the Che approximation presented in the last section. This yields the popularity of objects at the second layer, $q'(n) = q(n)(1-h(n))$. We can reapply the Che approximation with $q'$ to derive the $h'(n)$. The overall hit rate is then $ \sum_n q(n)(h(n)+h'(n)-h(n)h'(n))/ \sum_n q(n)$.

\begin{figure}[ht]
\centerline{\epsfig{figure=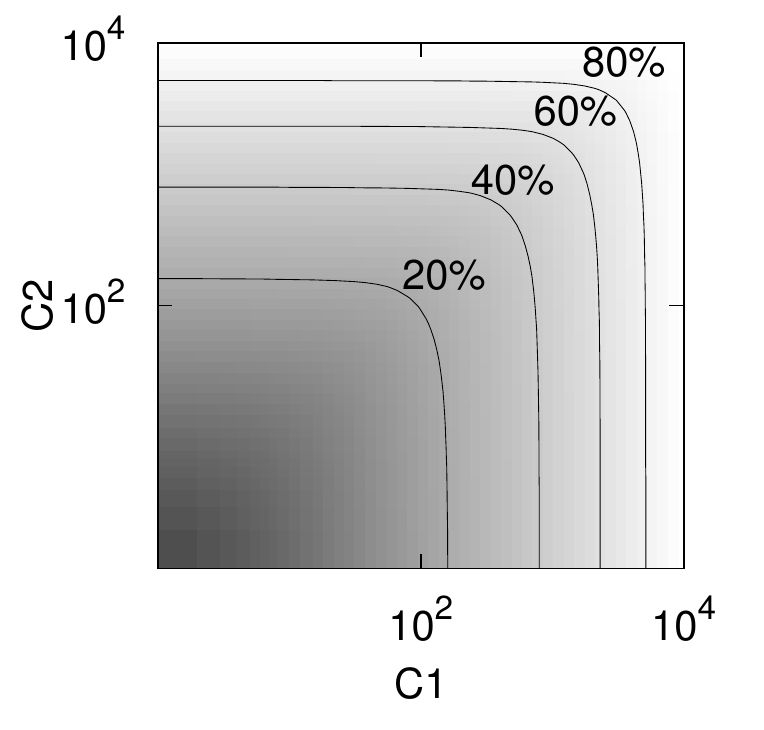,height=50mm,width=50mm}
\hspace{-8mm}\epsfig{figure=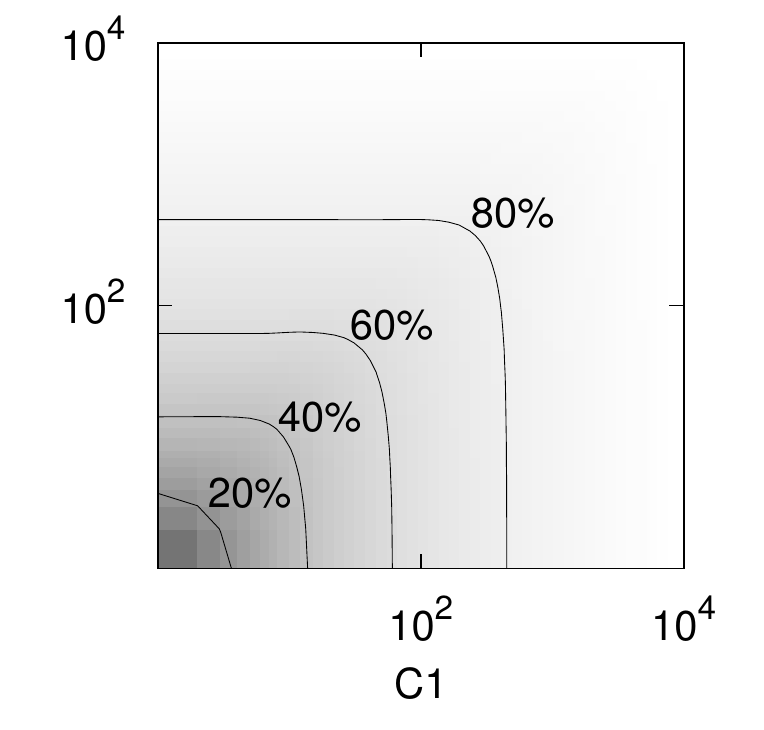,height=50mm,width=50mm}}
\vspace{-5mm}
    \caption{%
        Hit rate(\%) as a function of cache size in layers 1 and 2: left, $\alpha=.8, N=10^4$; right, $\alpha=1.2, N=10^4$.
     }%
     \label{fig:hit-hierarchy}
     \vspace{-3mm}

\end{figure}

Figure \ref{fig:hit-hierarchy} shows the hit rate as a function of $C_1$ and $C_2$ in the form of contour plots for a population of $10^4$ objects and Zipf popularity distributions with parameters $\alpha=.8$ and $\alpha=1.2$, respectively. The darker the shade of grey, the lower the hit rate. The contours show pairs ($C_1, C_2$) for which the overall hit rate is equal to the indicated percentage. The figures are nearly symmetrical about the diagonal. The iso-hit-rate contours reveal that overall performance depends essentially on the sum $C_1+C_2$ although the hit rate does decrease somewhat as $C_1$ and $C_2$ tend to equality. This occurs because both layers then store a larger number of the same, most popular objects.  

\subsection{Discussion}
\label{sec:discussion1}
Note first that the hit rate does not depend on traffic intensity. The proportional bandwidth reduction between layers 1 and 2 is thus independent of where layer 1 caches are located while overall memory cost increases as they are situated in more numerous routers or concentrators closer to users. 

Caches placed within routers (as proposed in \cite{JSTP09}) will likely be relatively small for reasons of cost and performance \cite{Perino2011}. In terms of objects, the considered population of $10^4$ (like the VoD catalogue assumed in Table \ref{tab:characteristics}) is more amenable to such caching than content with a population of $10^8$ objects (typical of UGC). We return to the contrast between VoD and UGC (and web and file sharing)  in the next section.

For both an isolated cache and a two-layer hierarchy, the Zipf law parameter $\alpha$ has a significant impact on the cache size required to attain a target hit rate (and consequent reduction in network bandwidth requirements).


\section{Caching the content mix}
\label{sec:mix}
We evaluate cache performance when requests are for objects of the 4 types identified in Section \ref{sec:traffic}.
The cache does not discriminate according to type but applies LFU or LRU replacement on the basis of the overall object request process. We assume objects of the same type have the same size.

\subsection{LFU performance}
For each unit of traffic demand, let $q_i(n) = r_i/n^{\alpha_i}$ be the arrival rate of requests for the rank $n$ object of type $i$ (notation is listed in Table \ref{tab:notation}). For definiteness, we suppose this is a normalized rate such that $\sum_i \sum_n q_i(n) \theta_i = 1$. Thus, by Little's formula, $r_i=p_i /(\theta_i \sum_{n=1}^{N_i} n^{-\alpha_i}) $. 

\begin{figure}[t]
\vspace{-2mm}
\centering
\includegraphics[scale=.6]{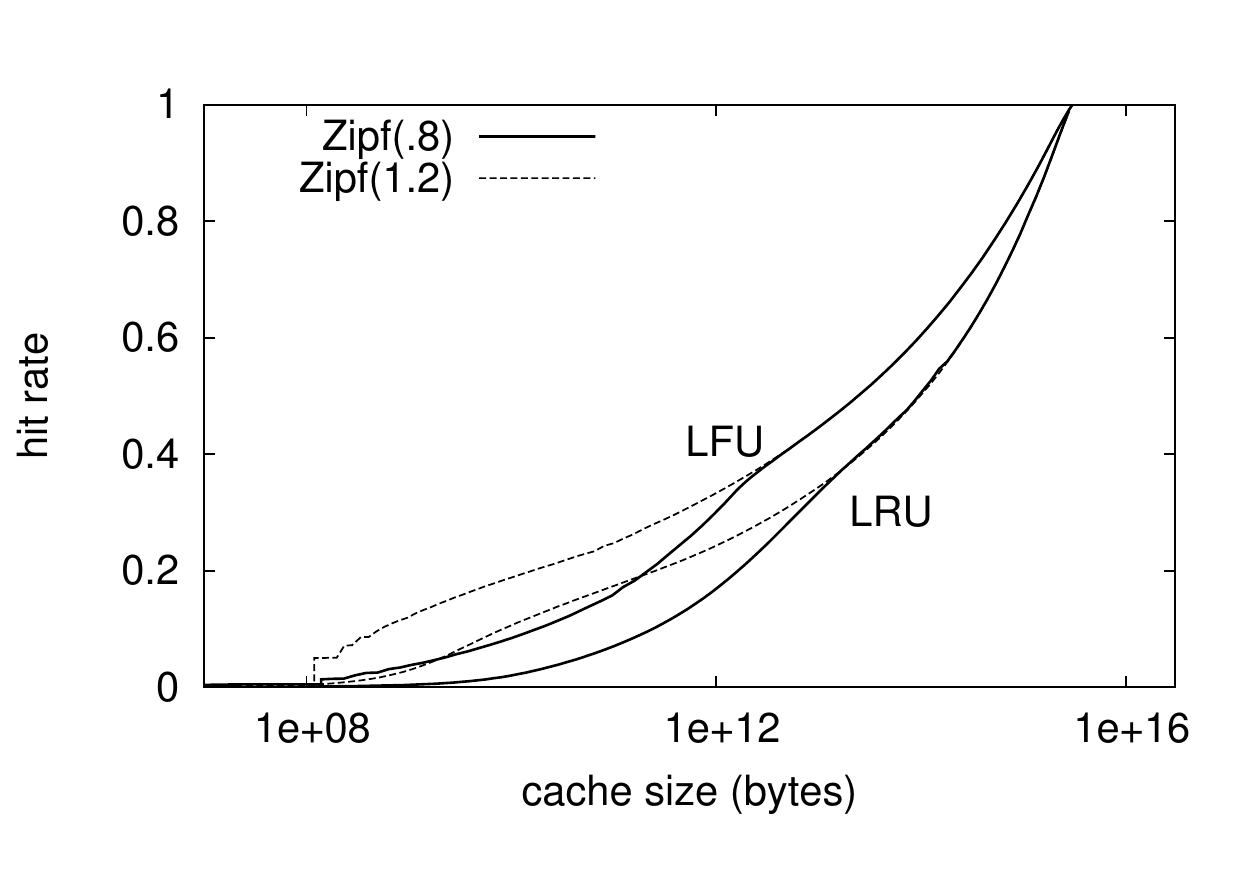}
\vspace{-4mm}
\caption{Overall hit rate for traffic mix:  2011 traffic, LFU and LRU replacement, Zipf(.8) and Zipf(1.2) for VoD content.}
\label{fig:mix-LFU-LRU}
\vspace{-3mm}

\end{figure}

Let $\mathsf{size}(x)$ be the cache size needed to store all objects having an arrival rate greater than $x$ assuming LFU replacement:
\begin{equation}
\mathsf{size}(x) =  \sum_i \theta_i  \gamma_i(x),
\label{eq:size}
\end{equation}
where $\gamma_i(x) = \min(N_i, \lfloor (r_i/x)^{1/\alpha_i}\rfloor)$.
Let $\mathsf{hit}(x)$ be the hit rate for a cache of this size:
\begin{equation}
\mathsf{hit}(x) = \sum_i \sum_{n=1}^{\gamma_i(x)} q_i(n)\theta_i.
\label{eq:hit}
\end{equation}
Relations (\ref{eq:size}) and (\ref{eq:hit}) define a parametric plot of hit rate against cache size. 
Figure \ref{fig:mix-LFU-LRU} illustrates this relation for the data of Table \ref{tab:characteristics} for year 2011. We defer comments to the next section where the results for LRU replacement are derived.

\subsection{LRU performance}
The Che approximation introduced in Section \ref{sec:che} can be adapted as follows for mixed content. We assume there is a parameter $T_C$ such that the hit rate $h_i(n)$ of the rank $n$ object of type $i$ when cache size is $C$ bytes is given by $1-e^{-q_i(n)T_C}$. Reasoning as above for equation (\ref{eq:che}), $T_C$ is the root of the equation:
\begin{equation}
C = \sum_{i=1}^4 \sum_{n=1}^{N(i)} (1-e^{-q_i(n)T_C}) \theta_i.
\label{eq:che4}
\end{equation}
The overall hit rate plotted in Figure \ref{fig:mix-LFU-LRU} is the sum $\sum_i \sum_n q_i(n) h_i(n)$. 

Note that (\ref{eq:che4}) would give precisely the same results had we identified chunks instead of entire objects under the assumption that each chunk inherits the popularity of its parent object. This justifies our choice to model content in terms of objects, even if some applications and/or network architectures would actually manage content at chunk level. Note also that to assume constant object size only impacts the performance evaluation of relatively small caches. Otherwise, the central limit theorem ensures our estimates are accurate, even though the true object size distribution may have a large variance.

The results in Figure  \ref{fig:mix-LFU-LRU} confirm the relative loss in hit rate of LRU compared to LFU observed for homogeneous content (cf. Fig \ref{fig:LFU-LRU}).  This adds credence to the accuracy of the adapted Che approximation which it is not practical to check by simulation here in view of the huge object populations considered. The impact of the Zipf exponent for VoD popularity is apparent for cache sizes where this type of content occupies a significant part of the cache.

\begin{figure}[t]
\centering
\includegraphics[scale=.6]{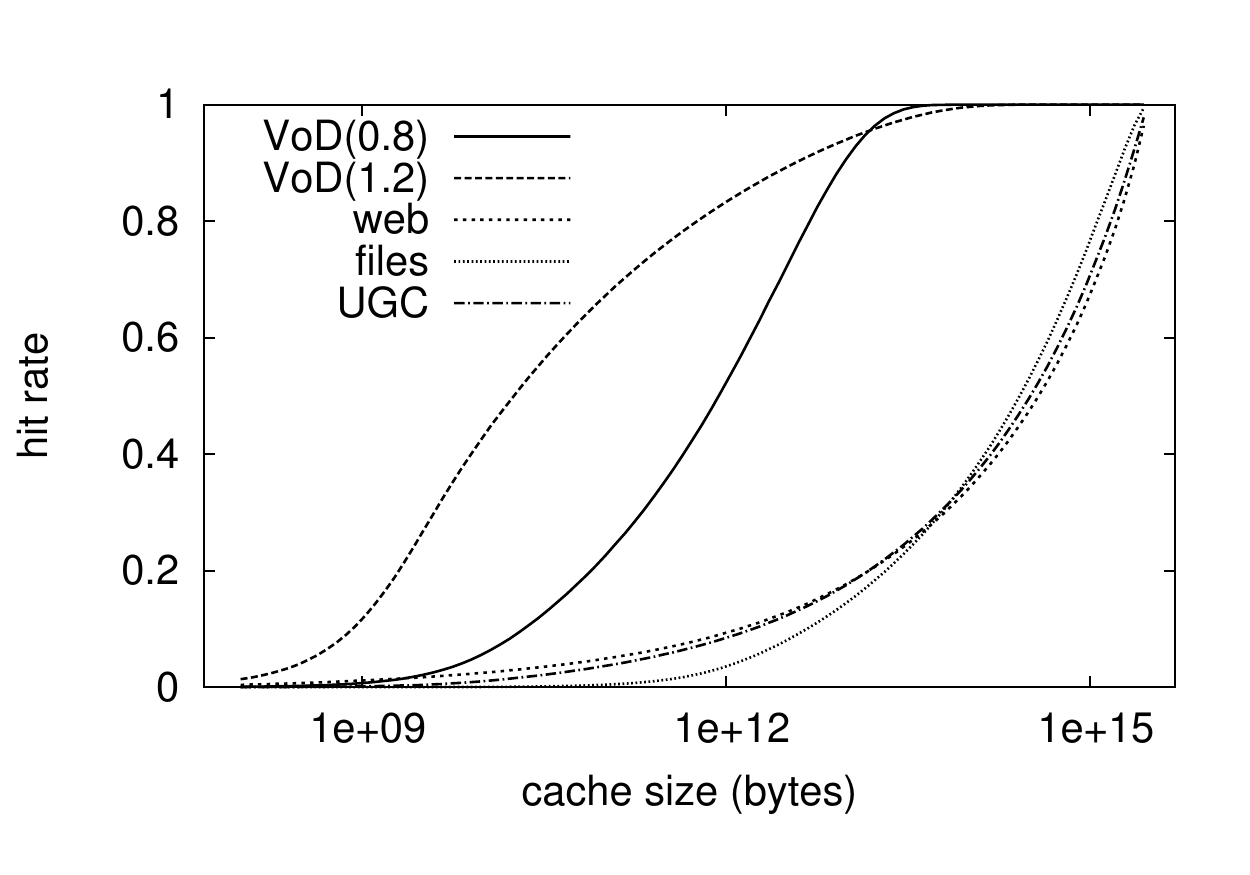}
\vspace{-4mm}
\caption{Hit rates for traffic types:  2011 traffic, Zipf(.8) and Zipf(1.2) VoD popularity (plots coincide for web, file sharing and UGC).}
\label{fig:hits4}
\vspace{-3mm}

\end{figure} 

Figure \ref{fig:hits4} shows how the hit rate for each of the 4 types evolves with cache size. The figure superposes results for Zipf(.8) and Zipf(1.2) VoD popularity. The hit rates for the other types of content only depend marginally on this and their plots coincide.  The high hit rates for VoD arise thanks mainly to the limited catalogue size for this type of content though, as expected (cf. Fig \ref{fig:LFU-LRU}), performance for small cache sizes is significantly better with the higher Zipf exponent. 

\begin{figure}[t]
\centering
\includegraphics[scale=.6]{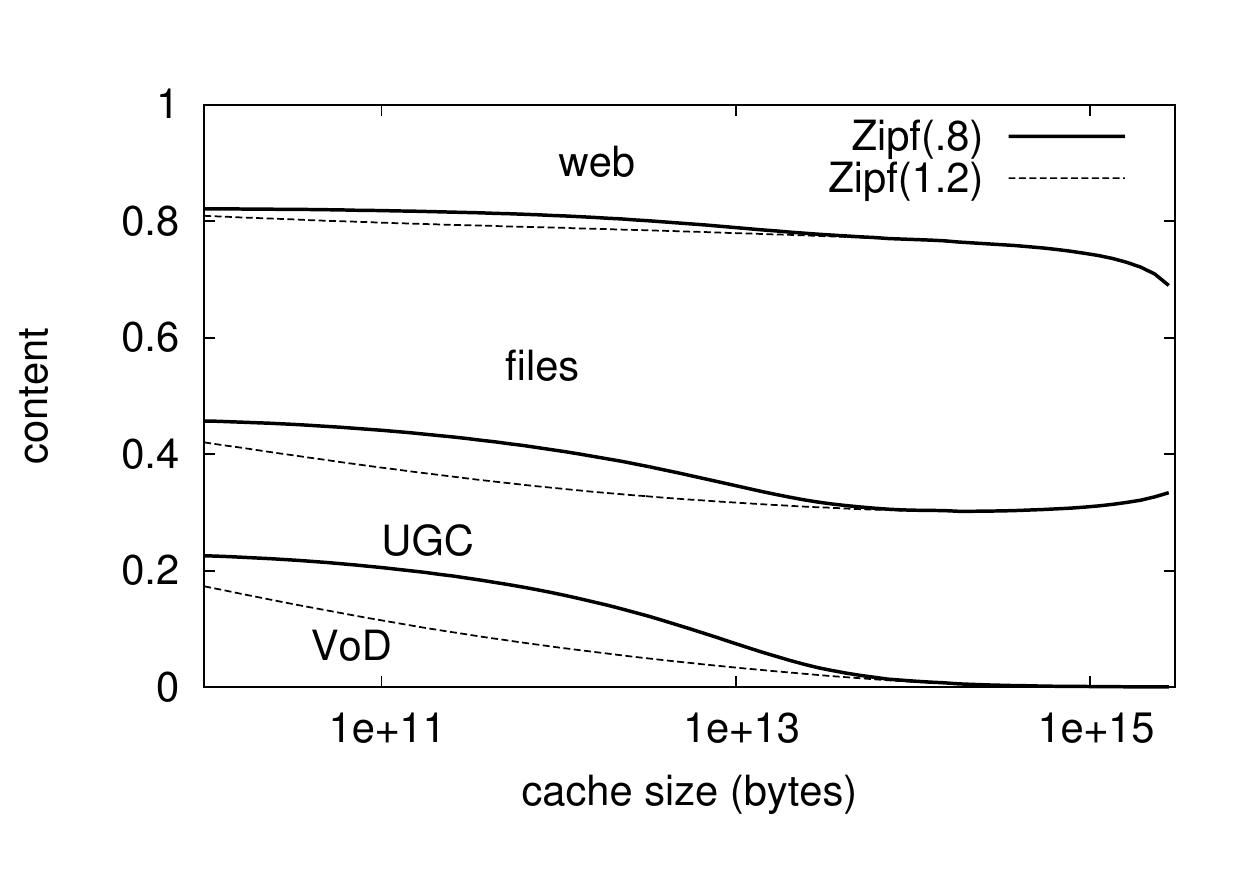}
\vspace{-4mm}
\caption{Cache utilization as function of cache size:  2011 traffic, Zipf(.8) and Zipf(1.2) VoD popularity.}
\label{fig:contents}
\vspace{-3mm}

\end{figure}

Figure \ref{fig:contents} shows how the cache is shared between the four types of content. When the cache is small, memory is occupied in proportion to the traffic shares $p_i$ since the hit rate for all objects is then very low. As the cache size increases the main observation is that the proportion of memory occupied by VoD content diminishes becoming negligible for a cache of 10 terabytes. 
The share occupied by VoD content is smaller for Zipf(1.2) popularity since requests then tend to be concentrated on a smaller number of very popular video objects.

\subsection{A two-layer cache hierarchy}

Proceeding as above for homogeneous content, we can calculate the hit rate for each type in the considered content mix. We evaluate the hit rates at layer 1 as above and deduce modified request rates at layer 2. Reapplying the modified Che approximation yields second layer hit rates and consequently the overall hit rate.

\begin{figure}[ht]
\centerline{\epsfig{figure=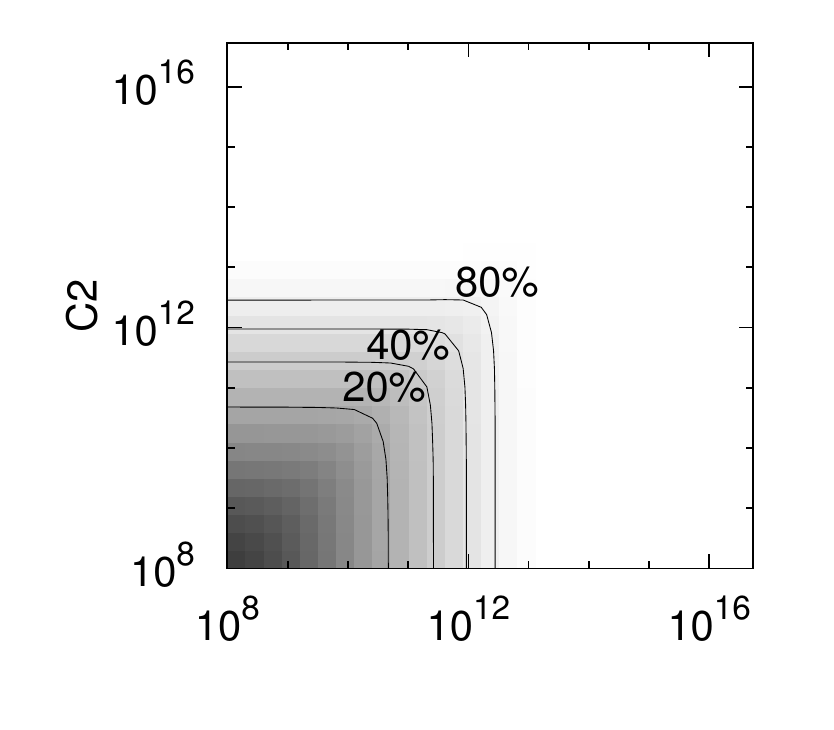,height=45mm,width=52mm}
\hspace{-10mm}\epsfig{figure=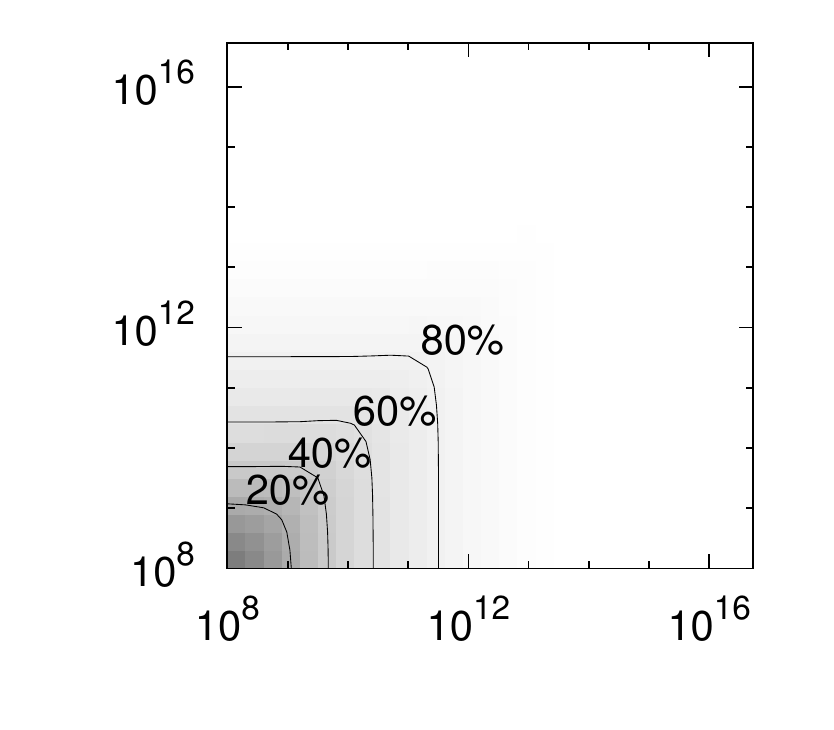,height=45mm,width=52mm}}
\vspace{-5mm}
\centerline{\epsfig{figure=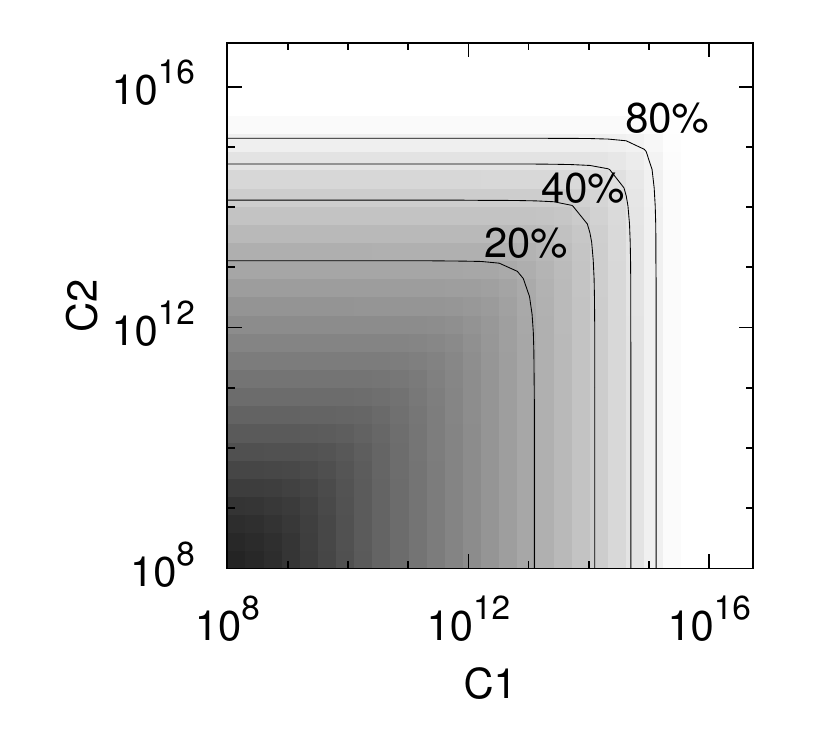,height=45mm,width=52mm}
\hspace{-10mm}\epsfig{figure=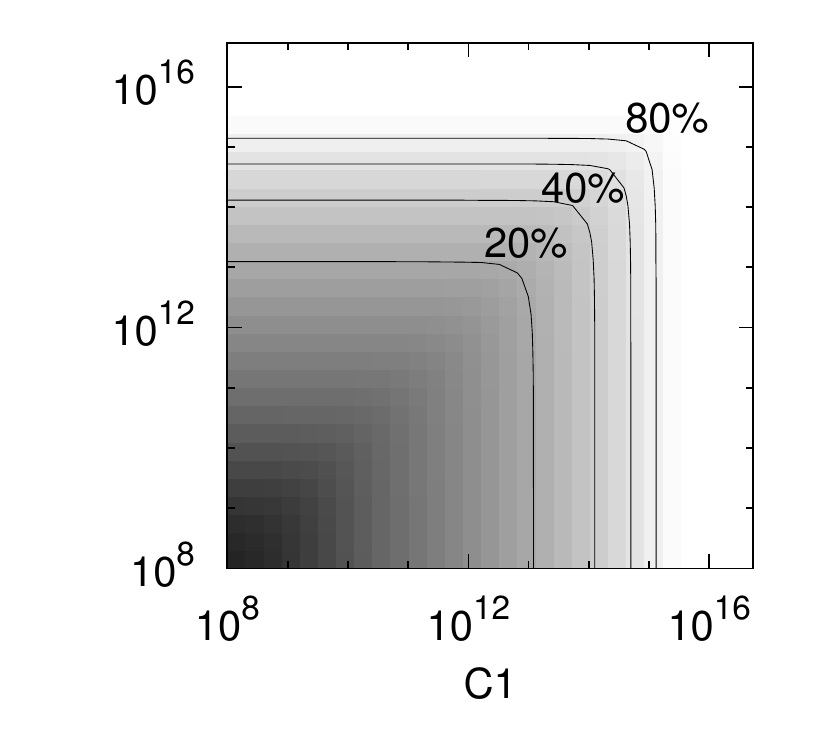,height=45mm,width=52mm}}
   \label{fig:contours4}
  \vspace{-2mm}
    \caption{ Hit rate (\%) as a function of cache size in layers 1 and 2: top left, VoD, Zipf(.8); top right, VoD,Zipf(1.2), bottom left, UGC, Zipf(.8), bottom right, UGC, Zipf(1.2).}%
    \label{fig:contours41}
\end{figure}

It turns out that UGC, file sharing and web types all have roughly the same behaviour, as might be expected from the results of Figure \ref{fig:hits4}. We therefore only plot in Figure \ref{fig:contours41} the overall hit rate for VoD and UGC content, respectively. The figure reveals contrasting performance. VoD content achieves a high hit rate with a relatively small  layer 1 cache, of one terabyte say, and has little need for  layer 2. On the other hand, UGC content only achieves a similar hit rate with  a total of around 100 terabytes, distributed between $C_1$ and $C_2$. The VoD Zipf law parameter $\alpha$ has a strong impact on the VoD hit rate but not on that of UGC.

\subsection{Discussion}
\label{sec:discussion2}
The above results confirm the observation in Section \ref{sec:discussion1}: a layer 1 cache small enough to be incorporated in an access router (e.g., a memory of 1 TB) is effective for VoD but not for the other types of content. Significant bandwidth reduction for the latter requires a much bigger cache that we would expect to take the form of specialized storage devices located in the core at layer 2. 

In Table \ref{tab:savings} we evaluate the bandwidth saving realized by layer 1 caches of size 1 TB and a layer 2 cache of size 100 TB. 
We evaluate the savings upstream of each layer assuming the first layer is either shared by all content or dedicated to VoD only. We perform this evaluation for the traffic mixes for 2011 and 2015 given in Table \ref{tab:characteristics}.

\begin{table}[t]
\begin{center}
\begin{tabular}{c | c | c | c | c }
\hline
 & Zipf VoD &  layer 1  &  \multicolumn{2}{c}{bandwidth reduction (\%)}\\
  & ($\alpha$) & cache & layer 1 & layers 1 \& 2 \\
   \hline
2011   & 0.8 & shared & 17 & 50 \\
  &   & VoD & 23  &  58 \\
   & 1.2 & shared &  24 & 50  \\
& & VoD & 23 & 58  \\ 
   \hline
2015   & 0.8 & shared & 27 & 59 \\
  &   & VoD & 37  &  61 \\
   & 1.2 & shared &  36 & 59  \\
& & VoD & 37 & 61  \\ 
 \hline
\end{tabular}

\caption{Bandwidth savings for $C_1=$ 1 TB and $C_2=$ 100 TB.}
\label{tab:savings}
\vspace{-8mm}

\end{center}
\end{table}

If VoD has Zipf(.8) popularity, dedicating layer 1 to VoD would be significantly more efficient. The evaluations reveal little to no gain for a VoD Zipf exponent of 1.2, however, since the layer 1 hit rate for VoD is high in both cases. Of course, if VoD turns out to have such accentuated popularity, one could reduce the size of layer 1 caches.


\section{Conclusion}
The bandwidth memory tradeoff realized by network caching depends significantly on the characteristics of the four main types of content: web, file sharing, UGC and VoD. This paper has evaluated the performance of a simple two-layer cache hierarchy under realistic traffic assumptions.

To significantly reduce bandwidth requirements for the first three types of content, each corresponding to a volume of around 1 petabyte and having a Zipf popularity law with a low exponent, namely 0.8, a large cache of at least 100 TB is needed. It appears likely that such capacity could only be economically provided in the network core, at what we have called layer 2.

On the other hand, VoD content is characterized by a relatively small catalogue totaling only 1 TB and, in view of its high and increasing traffic share, could advantageously be cached within the lower layer of access router content stores. Hit rate comparisons suggest it may be preferable to dedicate the layer 1 caches to this type of traffic, if possible, rather than caching all content indiscriminately.

The presented analysis is based on estimated characteristics that it would clearly be desirable to make more reliable. It is particularly important to establish the popularity law of VoD services since performance depends critically on this. 

Evaluations have been performed using what we termed the Che approximation. We numerically validated the accuracy of this approximation, especially for the large populations and cache sizes that are relevant for the present study. However, it largely remains to discover the mathematical arguments that explain this precision.

\providecommand{\bysame}{\leavevmode\hbox to3em{\hrulefill}\thinspace}
\providecommand{\MR}{\relax\ifhmode\unskip\space\fi MR }
\providecommand{\MRhref}[2]{%
  \href{http://www.ams.org/mathscinet-getitem?mr=#1}{#2}
}
\providecommand{\href}[2]{#2}

\end{document}

%% file: network.pdf_t
\begin{picture}(0,0)%
\includegraphics{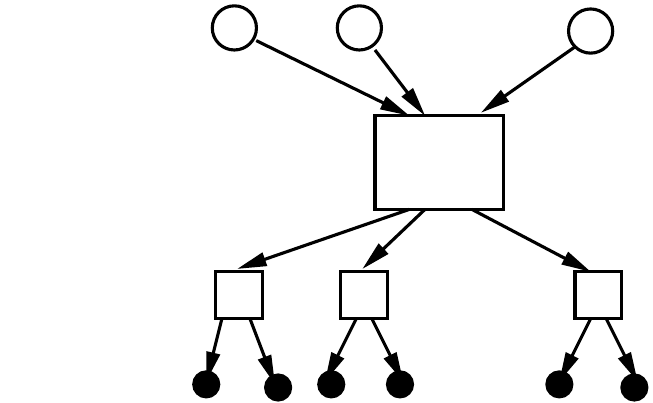}%
\end{picture}%
\setlength{\unitlength}{3947sp}%
\begingroup\makeatletter\ifx\SetFigFont\undefined%
\gdef\SetFigFont#1#2#3#4#5{%
  \reset@font\fontsize{#1}{#2pt}%
  \fontfamily{#3}\fontseries{#4}\fontshape{#5}%
  \selectfont}%
\fi\endgroup%
\begin{picture}(3120,1920)(2461,-1740)
\put(2476, 45){\makebox(0,0)[lb]{\smash{{\SetFigFont{12}{14.4}{\familydefault}{\mddefault}{\updefault}{\color[rgb]{0,0,0}sources}%
}}}}
\put(4696, 29){\makebox(0,0)[lb]{\smash{{\SetFigFont{20}{24.0}{\familydefault}{\mddefault}{\updefault}{\color[rgb]{0,0,0}.}%
}}}}
\put(4846, 29){\makebox(0,0)[lb]{\smash{{\SetFigFont{20}{24.0}{\familydefault}{\mddefault}{\updefault}{\color[rgb]{0,0,0}.}%
}}}}
\put(4561,-1246){\makebox(0,0)[lb]{\smash{{\SetFigFont{20}{24.0}{\familydefault}{\mddefault}{\updefault}{\color[rgb]{0,0,0}.}%
}}}}
\put(4711,-1246){\makebox(0,0)[lb]{\smash{{\SetFigFont{20}{24.0}{\familydefault}{\mddefault}{\updefault}{\color[rgb]{0,0,0}.}%
}}}}
\put(4861,-1246){\makebox(0,0)[lb]{\smash{{\SetFigFont{20}{24.0}{\familydefault}{\mddefault}{\updefault}{\color[rgb]{0,0,0}.}%
}}}}
\put(2491,-661){\makebox(0,0)[lb]{\smash{{\SetFigFont{12}{14.4}{\familydefault}{\mddefault}{\updefault}{\color[rgb]{0,0,0}layer 2}%
}}}}
\put(2506,-1291){\makebox(0,0)[lb]{\smash{{\SetFigFont{12}{14.4}{\familydefault}{\mddefault}{\updefault}{\color[rgb]{0,0,0}layer 1}%
}}}}
\put(2506,-1696){\makebox(0,0)[lb]{\smash{{\SetFigFont{12}{14.4}{\familydefault}{\mddefault}{\updefault}{\color[rgb]{0,0,0}users}%
}}}}
\put(4546, 29){\makebox(0,0)[lb]{\smash{{\SetFigFont{20}{24.0}{\familydefault}{\mddefault}{\updefault}{\color[rgb]{0,0,0}.}%
}}}}
\end{picture}%

%% file: nomen.bbl
\begin{thebibliography}{10}

\bibitem{Breslau99}
L.~Breslau, Pei Cao, Li~Fan, G.~Phillips, and S.~Shenker, \emph{Web caching and
  zipf-like distributions: evidence and implications}, INFOCOM'99, vol.~1,
  March 1999, pp.~126 --134.

\bibitem{Carlinet11}
Y.~Carlinet, B.~Kauffmann, P.~Olivier, and A.~Simonian, \emph{Trace-based
  analysis for caching multimedia services}, Orange labs technical report,
  2011.

\bibitem{CGM11}
G.~Carofiglio, M.~Gallo, and L.~Muscariello, \emph{Bandwidth and storage
  sharing performance in information centric networking}, ACM Sigcomm workshop
  on ICN, 2011.

\bibitem{Cha07}
Meeyoung Cha, Haewoon Kwak, Pablo Rodriguez, Yong-Yeol Ahn, and Sue Moon,
  \emph{I tube, you tube, everybody tubes: analyzing the world's largest user
  generated content video system}, Proceedings of the 7th ACM SIGCOMM
  conference on Internet measurement (New York, NY, USA), IMC'07, ACM, 2007,
  pp.~1--14.

\bibitem{CTW02}
Hao Che, Ye~Tung, and Zhijun Wang, \emph{Hierarchical web caching systems:
  modeling, design and experimental results}, {IEEE JSAC} \textbf{20} (2002),
  no.~7, 1305--1314.

\bibitem{CVNI}
Cisco, \emph{Cisco visual networking index: Forecast and methodology,
  2010-2015}, White paper, 2011.

\bibitem{FGT92}
P.~Flajolet, D.~Gardy, and L.~Thimonier, \emph{Birthday paradox, coupon
  collectors, caching algorithms and self-organizing search}, Discrete Appllied
  Mathematics \textbf{39} (1992), 207--229.

\bibitem{Gill07}
Phillipa Gill, Martin Arlitt, Zongpeng Li, and Anirban Mahanti, \emph{Youtube
  traffic characterization: a view from the edge}, Proceedings of the 7th ACM
  SIGCOMM conference on Internet measurement (New York, NY, USA), IMC'07, ACM,
  2007, pp.~15--28.

\bibitem{JSTP09}
V.~Jacobson, D.~Smetters, J.~Thornton, M.~Plass, N.~Briggs, and R.~Braynard,
  \emph{Networking named content}, {CoNext 2009}, 2009.

\bibitem{JKR05}
P.~Jelenkovic, X.~Kang, and A.~Radovanovic, \emph{Near optimality of the
  discrete persistent access caching algorithm}, International Conference on
  Analysis of Algorithms,, 2005.

\bibitem{Jelenkovic99}
P.~R. Jelenkovic, \emph{Approximation of the move-to-front search cost
  distribution and least-recently-used caching fault probabilities}, Annals of
  Applied Probability \textbf{9} (1999), no.~2, 430--464.

\bibitem{JR08}
P.~R. Jelenkovic and A.~Radovanovic, \emph{The persistent-access-caching
  algorithm}, Random Structures and Algorithms \textbf{33} (2008), no.~2,
  219--251.

\bibitem{Zhou11}
J.Zhou, Y.~Li, K.~Adhikari, and Z-L. Zhang, \emph{Counting youtube videos via
  random prefix sampling}, {Proceedings of IMC'07}, 2011.

\bibitem{LSS04}
N.~Laoutaris, S.~Syntila, and I.~Stavrakakis, \emph{Meta algorithms for
  hierarchical web caches}, Performance, Computing, and Communications, 2004
  IEEE International Conference on, 2004, pp.~445 -- 452.

\bibitem{Mahanti00}
A.~Mahanti, C.~Williamson, and D.~Eager, \emph{Traffic analysis of a web proxy
  caching hierarchy}, {IEEE Network} (2000), 16--23.

\bibitem{Perino2011}
D.~Perino and M.~Varvello, \emph{A reality check for content centric
  networking}, ACM Sigcomm workshop on ICN, 2011.

\bibitem{Psaras2011}
I.~Psaras, R.~G. Clegg, R.~Landa, W.~K. Chai, and G.~Pavlou, \emph{Modelling
  and evaluation of {CCN} caching trees}, IFIP Networking 2011, 2011.

\bibitem{Williamson02}
Carey Williamson, \emph{On filter effects in web caching hierarchies}, ACM
  Trans. Internet Technol. \textbf{2} (2002), 47--77.

\bibitem{Yu06}
Hongliang Yu, Dongdong Zheng, Ben~Y. Zhao, and Weimin Zheng,
  \emph{Understanding user behavior in large-scale video-on-demand systems},
  SIGOPS Oper. Syst. Rev. \textbf{40} (2006), 333--344.

\end{thebibliography}
